\documentclass[journal]{IEEEtran}
\pdfoutput=1
\ifCLASSINFOpdf
\else
   \usepackage[dvips]{graphicx}
\fi
\usepackage{url}
\usepackage{subfig} 
\usepackage{algorithm,algorithmic}
\usepackage{graphicx}
\usepackage{amsmath}
\usepackage{amssymb}
\usepackage{cite}
\usepackage{tabularx}
\usepackage{booktabs}
\makeatletter
\let\NAT@parse\undefined
\makeatother
\usepackage{hyperref}  
\hypersetup{hypertex=true,
	colorlinks=true,
	linkcolor=blue,
	anchorcolor=blue,
	citecolor=magenta}
\begin{document}
\title{Joint DOA and Non-circular Phase Estimation of Non-circular Signals for Antenna Arrays: Block Sparse Bayesian Learning Method }

\author{Zihan Shen, Jiaqi Li\textsuperscript{*}, Xudong Dong and Xiaofei Zhang
\thanks{This work was supported in part by National Natural Science Foundation of China under Grants 62371225, 62371227. (Corresponding author: Jiaqi Li)}
\thanks{Zihan Shen, Jiaqi Li, and Xiaofei Zhang are with the Key Laboratory of Dynamic Cognitive System of
Electromagnetic Spectrum Space, Nanjing University of Aeronautics and Astronautics,  Nanjing 211106, China (e-mail: shenzihan@nuaa.edu.cn, lijiaqi2000@outlook.com, zhangxiaofei@nuaa.edu.cn).}
\thanks{Xudong Dong is with the College of Communication Engineering, Hangzhou Dianzi University, Hangzhou 310018, China (e-mail: dxd@hdu.edu.cn).}}

\maketitle

\begin{abstract}
This letter proposes a block sparse Bayesian learning (BSBL) algorithm of non-circular (NC) signals for direction-of-arrival (DOA) estimation, which is suitable for arbitrary  unknown NC phases. The block sparse NC signal representation model is constructed through a permutation strategy, capturing the available intra-block structure information to enhance recovery performance. After that, we create the sparse probability model and derive the cost function under BSBL framework. Finally, the fast marginal likelihood maximum (FMLM) algorithm is introduced, enabling the rapid implementation of signal recovery by the addition and removal of basis functions. Simulation results demonstrate the effectiveness and the superior performance of our proposed method.
\end{abstract}


\IEEEpeerreviewmaketitle

\section{Introduction}
\parskip=0pt
\IEEEPARstart{C}ONVENTIONAL subspace-based algorithms, such as multiple signal classification (MUSIC) \cite{schmidt1986multiple}, estimation of signal parameters via rotational invariance techniques (ESPRIT) \cite{roy1986esprit}, and their variants like root-MUSIC \cite{10458936} and TLS-ESPRIT \cite{10663219}, are widely used for direction-of-arrival (DOA) estimation due to their high resolution. However, their performance critically depends on the number of snapshots and array elements \cite{shen2016underdetermined}.

Emerging algorithms based on sparse signal representation (SSR) models, such as \(\ell_p\)-norm (\(0 < p \leq 1\)) methods and greedy algorithms, frame DOA estimation as a sparse reconstruction problem for high accuracy \cite{niu2023doa, song2023fast}. Compared with the aforementioned algorithms, sparse Bayesian learning (SBL) achieves the global minimum at the sparsest solution with fewer local minima \cite{10065562,li2024enhanced,9540245}. Zhang et al. enhances the SBL with the block sparse Bayesian learning (BSBL) model, incorporating a block correlation matrix for improved accuracy \cite{6415293}. The multiple measurement vector (MMV) model is transformed into a single measurement vector (SMV) model, culminating in the development of the T-SBL algorithm \cite{5887383}.


Utilizing waveform characteristics can enhance DOA estimation performance, particularly with non-circular (NC) signals like multiple amplitude shift key (MASK) and binary phase shift key (BPSK), which exhibit unique properties including non-zero pseudo-covariance \cite{10124337, chen2015doa, wang2017non}. Several studies have proposed SBL algorithms for estimating NC signals \cite{zhang2020efficient, pan2024co}. However, these algorithms often operate under the assumption of zero and identical NC phases, which is impractical in real-world scenarios. Yuan et al. \cite{10038549} suggests that block sparsity can address this issue, but their method fails to accurately capture block sparsity or estimate NC phases.

This letter proposes an efficient DOA estimation algorithm of NC signals, enabling simultaneous recovery for DOA and NC phases within the BSBL framework. This algorithm constructs a block sparse model via a customized permutation matrix and then formulates a cost function using SBL and Type-II maximum likelihood \cite{6415293}. Finally the desired parameters are estimated through fast marginal likelihood maximization (FMLM). Key contributions summarized as follows:
\begin{enumerate}  
	\item We correct the defective SSR based NC signals model proposed in \cite{10038549} and construct an accurate one by our proposed permutation strategy.
	\item To recover DOA and unknown NC phases simultaneously, the block sparse Bayesian inference is leveraged and validated through simulations. 
	\item To reduce computational complexity, the FMLM algorithm is introduced, accelerating the iterative process meanwhile benefiting from block sparsity and NC signal properties.  
\end{enumerate}  

The notations used in this letter are described in Table \ref{table_notations}.
%
\vspace{-4.1mm}
\begin{table}[!h]
	\begin{center}
		\caption{Notational Conventions}
		\label{table_notations}
		\scriptsize
		\begin{tabular}{m{0.25\linewidth} m{0.65\linewidth}} 
			\toprule[1.15pt]
			$a$, $\mathbf{A}$, $\mathbf{a}$, $\mathbb{A}$ & Scalar, matrix, vector, and set \\
			$\mathbf{A}^\mathrm{T}$, $\mathbf{A}^{*}$, $\mathbf{A}^\mathrm{H}$, $\mathbf{A}^{-1}$ & Transpose, conjugate, Hermitian-transpose and inversion of matrix $\mathbf{A}$\\
			$\mathbf{A}=\text{diag}(a)$, $\mathbf{A}=\text{blkdiag}(a)$ & Diagonal matrix and a block diagonal matrix with given elements\\
			$[\mathbf{A}]_{i,\cdot}$, $[\mathbf{A}]_{\cdot,j}$, $[\mathbf{A}]_{i,j}$ & The elements of the $i$-th row, the elements of the $j$-th column, and the element in the $i$-th row and $j$-th column.\\
			$\mathbf{I}$, $\mathbf{O}$ & An identity matrix and a null matrix\\
			$\operatorname{Re}\{\mathbf{A}\}$, $\operatorname{Im}\{\mathbf{A}\}$ & The real part and the imaginary part of the matrix\\
			$|\mathbf{A}|$, $\Vert\mathbf{a}\Vert_2$, $\Vert\mathbf{A}\Vert_\mathrm{F}$ & The determinant of a matrix, the $\ell_2$-norm of a vector, and the Frobenius norm of a matrix\\
			$\text{Tr}\{\mathbf{A}\}$, $\text{Toeplitz}(\cdot)$, $\mathrm{E}\{\cdot\}$ & The trace of a matrix, a Toeplitz matrix with given elements and the mathematical expectation\\
			$\mathrm{j}=\sqrt{-1}$,  $\mathcal{CN}(,)$ & The imaginary unit and the complex Gaussian distribution\\
			$\mathrm{log}(a)$, $\Omega$, \(\delta(a, b)\) & The natural logarithm with base \( e \), the region of interest and the Kronecker delta function\\
			\bottomrule[1.15pt]
		\end{tabular}
	\end{center}
\end{table}

\section{Signal model}
\subsection{NC signal model}
Considering $K$ far-field independent narrow-band signals impinging on a uniform linear array (ULA) with $M$ sensors \cite{10328457},
\begin{equation}
	\mathbf{z}(t)=\mathbf{A}\mathbf{s}(t)+\mathbf{n}(t), t=1,2,\cdots,L,
	\label{received signal_1}
\end{equation}
where $\mathbf{A}=[\mathbf{a}(\theta_1),\mathbf{a}(\theta_2),\cdots,\mathbf{a}(\theta_K)] \in \mathbb{C}^{M \times K}$ with $\mathbf{a}(\theta_k)=[1,e^{-\mathrm{j}\pi \mathrm{sin}\theta_k},\cdots,e^{-\mathrm{j}(M-1)\pi \mathrm{sin}\theta_k}]^{\mathrm{T}}, k=1,2,\cdots,K$ and $\theta_{k}$ represents the DOA of the $k$-th source, $L$ denotes the number of snapshots. 

Collecting $L$ snapshots, (\ref{received signal_1}) can be written as
\begin{equation}
	\mathbf{Z}=\mathbf{A}\mathbf{S}+\mathbf{N} \in  \mathbb{C}^{M \times L},
	\label{received signal}
\end{equation}
where $\mathbf{Z}=[\mathbf{z}(1),\cdots,\mathbf{z}(L)]$, $\mathbf{S}=\boldsymbol{\Psi}\mathbf{S}_{R} \in \mathbb{C}^{K \times L}$ is the strict NC signals \cite{10124337}, $\boldsymbol{\Psi}=\mathrm{diag}(e^{-j\phi_{1},\dots,-j\phi_{K}}) \in \mathbb{C}^{K \times K}$ denotes the maximum noncircularity rate of signals and $\mathbf{S}_{R}=[\mathbf{s}_{R}(1),\cdots,\mathbf{s}_{R}(L)]$ denotes the real part of signals. 

The augmented received signal based on NC properties \cite{10124337} is
\begin{equation}
	\small
	\begin{aligned}
		\mathbf{Y}&=\left[\begin{array}{l}
			\mathbf{Z}^{\mathrm{*}}\\
			\mathbf{Z}
		\end{array}
		\right]=\left[\begin{array}{cc}
		{\mathbf{A}}^{*} & \mathbf{O}\\
			\mathbf{O} & {\mathbf{A}}
		\end{array}
		\right]\left[\begin{array}{l}
			{\mathbf{S}}^{\mathbf{*}}\\
			{\mathbf{S}}
		\end{array}
		\right]+\left[\begin{array}{l}
			\mathbf{N}^{*}\\
			\mathbf{N}
		\end{array}
		\right]\\
		&=\mathbf{B}{\mathbf{S}_\text{nc}}+\mathbf{W} \in \mathbb{C}^{2M \times L},
	\end{aligned}
	\label{Y}
\end{equation}
where $\mathbf{B}=\text{blkdiag}({\mathbf{A}}^{*},{\mathbf{A}}) \in \mathbb{C}^{2M\times 2K}$, $\mathbf{S}_\text{nc}=[{\mathbf{S}}^{\mathrm{H}},{\mathbf{S}^{\mathrm{T}}}]^\mathrm{T}$, $\mathbf{W}=[\mathbf{N}^{\mathrm{H}},\mathbf{N}^{\mathrm{T}}]^\mathrm{T}$. 
\subsection{On-grid sparse signal representation model}
We only consider the SSR based model of NC signals under the on-grid condition. In this case for multiple sources, the region of interest (ROI) $\Omega$ is divided into a set of uniform discrete grid points (GPs) \cite{li2024enhanced}, denoted as $\mathcal{G} = \{\tilde{\theta}_n \mid n = 1, \cdots, N\}$, where $N \gg K$. If all the source DOAs $\theta_k$ lie on the GPs, i.e., $\{\theta_k\}_{k=1}^{K}  \subseteq \mathcal{G}$. $\mathbf{Y}$ represents the received signal of $\overline{\mathbf{S}}$, and can be expressed as

\begin{equation}
	\mathbf{Y} = \overline{\mathbf{B}}\,\overline{\mathbf{S}}+\mathbf{W},
\end{equation}
where $\overline{\mathbf{B}}=\text{blkdiag}(\overline{\mathbf{A}}^{\mathrm{*}},\overline{\mathbf{A}}) \in \mathbb{C}^{2M \times 2N}$, $\overline{\mathbf{A}}=[\mathbf{a}(\tilde{\theta}_1),\cdots,\mathbf{a}(\tilde{\theta}_N)] \in \mathbb{C}^{M \times N}$. $\overline{\mathbf{S}}=[\tilde{\mathbf{S}}^{\mathrm{H}},\tilde{\mathbf{S}}^{\mathrm{T}}]^{\mathrm{T}}$, $\tilde{\mathbf{S}}$ is a row-sparse matrix, expressed as
\begin{equation}
	[\tilde{\mathbf{S}}]_{{n,\cdot}}=	\begin{cases}
		[\mathbf{S}]_{k,\cdot}, & \text{if } \tilde{\theta}_n = \theta_k,  \\
		\mathbf{0}, & \text{otherwise}.
	\end{cases}
\end{equation}

\section{Proposed Algorithm}
\subsection{Accurate block sparse signal model}
\begin{figure}[!ht]
	\centerline{\includegraphics[width=0.7\columnwidth]{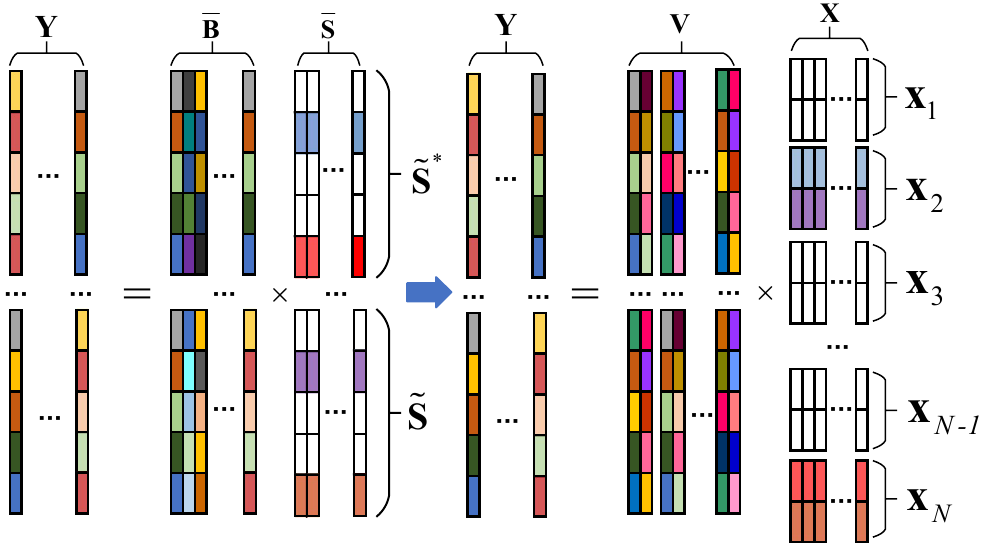}}
	\caption{Structure of block sparse signal model.}
	\label{structure}
\end{figure}

The signal $\overline{\mathbf{S}}$, as shown in the left part of Fig. \ref{structure}, does not yet possess block sparse characteristics but instead exhibits an element-wise sparsity pattern along its rows. Therefore, a permutation matrix $\mathbf{J}$ is needed to reconstruct the mainfold matrix, ensuring that the signal exhibits block-sparse characteristics.

Define $\mathbf{J} \in \mathbb{C}^{2N \times 2N}$ with its $(i,j)$-th entry expressed as
\begin{equation}
	\mathbf{J}_{i,j} = \delta(j, 2i-1) \cdot \mathbf{1}_{i \leq N} + \delta(j, 2i-2N) \cdot \mathbf{1}_{i > N},
\end{equation}
where \(\mathbf{1}_{\text{condition}}\) denotes the indicator function, which equals 1 if the condition is true, and 0 otherwise.

By reconstructing the array manifold matrix, the NC signal is converted into a block-structured sparse signal, as shown in the right part of Fig. \ref{structure},
\begin{equation}
	\mathbf{Y}=\overline{\mathbf{B}}\mathbf{J}\mathbf{X}+\mathbf{W}=\mathbf{V}\mathbf{X}+\mathbf{W},
	\label{V}
\end{equation}
where $\mathbf{V}=\overline{\mathbf{B}}\mathbf{J}$, and the block sparse signal, denoted by $\mathbf{X}$, consists of $N$ blocks $\mathbf{X}_{i} \in \mathbb{C}^{2 \times L}$, among which $K$ blocks are non-zero. $\tilde{\theta}_i$ denotes the angle of the $i$-th block. That is
\begin{equation}
	\mathbf{X}_{i}=
	\begin{cases}
		[e^{j\phi_{k}}\mathbf{S}_{R}^{\mathrm{T}},e^{-j\phi_{k}}\mathbf{S}_{R}^{\mathrm{T}}]^\mathrm{T}, & \text{if } \tilde{\theta}_i = \theta_{k},  \\
		\mathbf{O}, & \text{otherwise}.
	\end{cases}
\end{equation}

\subsection{BSBL framework for MMV model}
The $i$-th block signal $\mathbf{X}_{i}$ ($i=1,2,\cdots,N$) can be modeled as \cite{6415293}
\begin{equation}
	p(\mathbf{X}_{i};\{\gamma_{i},\mathbf{G}_{i}\})\sim\mathcal{CN}(\mathbf{X}_{i};\mathbf{O},\gamma_{i}\mathbf{G}_{i}),
\end{equation}
where $\mathbf{G}_i$ is a symmetric positive definite matrix that represents the correlation structure information within the signal block $\mathbf{X}_{i}$, $\gamma_{i}$ is a hyperparameter that controls the sparsity of the block signal $\mathbf{X}_{i}$. Assuming the inter-blocks of sparsity signals are mutually independent, the signal can be written as \cite{7536146}
\begin{equation}
	p(\mathbf{X};\boldsymbol{\Gamma}) = \prod_{i=1}^{N}p(\mathbf{X}_{i};\{\gamma_{i},\mathbf{G}_{i}\})\sim \mathcal{CN}(\mathbf{X};\mathbf{0},\mathbf{\Gamma}),
\end{equation}
where $\boldsymbol{\Gamma}=\mathrm{diag}(\gamma_{1}\mathbf{G}_1,\cdots,\gamma_{N}\mathbf{G}_N) \in \mathbb{C}^{2N \times 2N}$, and the weight vector $\boldsymbol{\gamma}=[\gamma_{1},\gamma_{2},\cdots,\gamma_{N}]^{\mathrm{T}}$.

$\mathbf{W}$ is an additive Gaussian white noise matrix, following a Gaussian distribution with a zero mean and variance $\beta^{-1}$ \cite{li2024enhanced},
\begin{equation}
	p(\mathbf{W})\sim\mathcal{CN}(\mathbf{O},\beta^{-1}\mathbf{I}).
\end{equation}

Therefore, the observation matrix $\mathbf{Y}$ follows a Gaussian distribution with a mean of $\mathbf{V}\mathbf{X}$ and variance $\beta^{-1}\mathbf{I}$ \cite{9540245},
\begin{equation}
	\begin{aligned}
		p(\mathbf{Y}|\mathbf{X};\beta)&\sim\mathcal{CN}(\mathbf{Y};\mathbf{V}\mathbf{X},\beta^{-1}\mathbf{I})\\
		&=(\pi\beta^{-1})^{-NL}\text{exp}\{-\beta\| \mathbf{Y}-\mathbf{V}\mathbf{X} \|_\mathrm{F}^{2}\}.
	\end{aligned}
\end{equation}

According to the law of total probability, the likelihood function can be derived as follows
\begin{equation}
	p(\mathbf{Y};\boldsymbol{\Gamma})=\int_{\mathbf{X}}p(\mathbf{Y}|\mathbf{X})p(\mathbf{X};\boldsymbol{\Gamma})\mathrm{d}\mathbf{X}.
	\label{gassuian}
\end{equation}

The integral in (\ref{gassuian}) can be interpreted as the convolution of two Gaussian functions, and the result still follows a Gaussian distribution \cite{7536146}
\begin{equation}
	p(\mathbf{Y}|{\boldsymbol{\Gamma}},\beta)\sim\mathcal{CN}(\mathbf{Y};\mathbf{O},\mathbf{C}),
	\label{C_1}
\end{equation}
where	$\mathbf{C}=\mathrm{E}\{\mathbf{Y}\mathbf{Y}^{\mathrm{H}}\}=\beta^{-1}\mathbf{I}+\mathbf{V}\boldsymbol{\Gamma}\mathbf{V}^{\mathrm{H}}$.

According to the Bayes rule, the posterior $p(\mathbf{X}|\mathbf{Y};\boldsymbol{\Gamma},\beta)$ can be expressed as \cite{7536146}
\begin{equation}
	\begin{aligned}
		& p(\mathbf{X}|\mathbf{Y};\boldsymbol{\Gamma},\beta)\propto p(\mathbf{Y}|\mathbf{X};\beta)p(\mathbf{X};\boldsymbol{\Gamma})\\
		& \propto \frac{\text{exp}\{-\text{Tr}[(\mathbf{X}-\boldsymbol{\mu})^{\mathrm{H}}\boldsymbol{\Sigma}^{-1}(\mathbf{X}-\boldsymbol{\mu})]\}}{(\pi^{N}|\boldsymbol{\Sigma}|)^{L}} \sim \mathcal{CN}(\boldsymbol{\mu},\boldsymbol{\Sigma}), 
	\end{aligned}
\end{equation}
where 
\begin{equation}
	\boldsymbol{\mu}=\beta\boldsymbol{\Sigma}\mathbf{V}^{\mathrm{H}}\mathbf{Y} \in \mathbb{C}^{2N \times L},
	\label{mu}
\end{equation}
\begin{equation}
	\boldsymbol{\Sigma}=(\boldsymbol{\Gamma}^{-1}+\beta\mathbf{V}^{\mathrm{H}}\mathbf{V})^{-1} \in \mathbb{C}^{2N \times 2N}.
	\label{sigma}
\end{equation}

To estimate the parameters $\{\gamma_{i}$,$\mathbf{G}_{i}\}$ and $\beta$, a Type-II maximum likelihood method can be used to obtain the cost function $\mathcal{L}(\boldsymbol{\Theta})$ \cite{6415293}, 
\begin{equation}
	\mathcal{L}(\boldsymbol{\Theta})=-2\mathrm{log}p(\mathbf{Y}|\boldsymbol{\Gamma},\beta)\propto L \text{log}|\mathbf{C}|+\text{Tr}(\mathbf{Y}^{\mathrm{H}}\mathbf{C}^{-1}\mathbf{Y}).
	\label{function}
\end{equation}
where $\boldsymbol{\Theta} \triangleq \{\beta,\{\gamma_{i},\mathbf{G}_{i}\}_{i=1}^{N}\}$ denotes a set of estimated parameters.

\subsection{Fast Marginal Likelihood Maximization}
In BSBL methods, parameter updates involve iterative algorithms with matrix inversion (\ref{sigma}), yielding $\mathcal{O}(8N^3)$ complexity. Using the Woodbury identity reduces it to $\mathcal{O}(8M^3)$, however, the computational cost is forbidden. To address this issue, the FMLM concept is introduced \cite{9540245}, where basis functions are iteratively added and removed until all non-zero weights are included.

$\mathbf{C}$ in (\ref{C_1}) can be rewritten as \cite{9540245}
\begin{equation}
	\begin{aligned}
		\mathbf{C}&=\beta^{-1}\mathbf{I}+ \sum_{n\neq 	i}\mathbf{V}_{n}\mathbf{L}_{n}\mathbf{V}_{n}^{\mathrm{H}}+     \mathbf{V}_{i}\mathbf{L}_{i}\mathbf{V}_{i}^{\mathrm{H}} \\
		&=\mathbf{C}_{-i}+\mathbf{V}_{i}\mathbf{L}_{i}\mathbf{V}_{i}^{\mathrm{H}}, i=1,2,\cdots, N,
		\label{C}
	\end{aligned}                      
\end{equation}
where $\mathbf{L}_{i} \triangleq \gamma_{i}\mathbf{G}_{i} \in \mathbb{C}^{2 \times 2}$ \cite{liu2014energy}, $\mathbf{C}_{-i}\triangleq\beta^{-1}\mathbf{I}+ \sum_{n\neq 	i}\mathbf{V}_{n}\mathbf{L}_{n}\mathbf{V}_{n}^{\mathrm{H}}$ is obtained by removing the $i$-th basic function from $\mathbf{C}$. $\mathbf{V}_{i} \in \mathbb{C}^{2M \times 2}$ denotes basis functions corresponding to the $i$-th signal block. 

%

Substituting the decomposition (\ref{C}) into the cost function (\ref{function}), resulting in
\begin{small}
	\begin{equation}
		\begin{aligned}
			\mathcal{L}(\boldsymbol{\Theta})&=L\mathrm{log}|\mathbf{C}_{-i}|+\text{Tr}[\mathbf{Y}^{\mathrm{H}}\mathbf{C}_{-i}^{-1}\mathbf{Y}]+L\mathrm{log}|\mathbf{I}+\mathbf{L}_{i}\mathbf{V}_{i}^{\mathrm{H}}\mathbf{C}_{-i}^{-1}\mathbf{V}_{i}|\\
			&-\text{Tr}\{\mathbf{Y}^{\mathrm{H}}\mathbf{C}_{-i}^{-1}\mathbf{V}_{i}(\mathbf{L}_{i}^{-1}+\mathbf{V}_{i}^{\mathrm{H}}\mathbf{C}_{-i}^{-1}\mathbf{V}_{i})^{-1}\mathbf{V}_{i}^{\mathrm{H}}\mathbf{C}_{-i}^{-1}\mathbf{Y}\}.
			\label{L}
		\end{aligned}
	\end{equation}
\end{small}

Let $\mathbf{U}_{i}\triangleq\mathbf{V}_{i}^{\mathrm{H}}\mathbf{C}_{-i}^{-1}\mathbf{V}_{i}$, $\mathbf{Q}_{i}\triangleq\mathbf{V}_{i}^{\mathrm{H}}\mathbf{C}_{-i}^{-1}\mathbf{Y}$, (\ref{L}) can be further simplified as

\begin{equation}
	\small
	\begin{aligned}
		\mathcal{L}(\boldsymbol{\Theta})&=L\mathrm{log}|\mathbf{C}_{-i}|+\text{Tr}[\mathbf{Y}^{\mathrm{H}}\mathbf{C}_{-i}^{-1}\mathbf{Y}]\\
		&+L\mathrm{log}|\mathbf{I}+\mathbf{L}_{i}\mathbf{U}_{i}|-\text{Tr}\{\mathbf{Q}_{i}^{\mathrm{H}}(\mathbf{L}_{i}^{-1}+\mathbf{U}_{i})^{-1}\mathbf{Q}_{i}\}\\
		&=\mathcal{L}(-i)+\mathcal{L}(i),
	\end{aligned}
\end{equation}
where $\mathcal{L}(i)\triangleq L\mathrm{log}|\mathbf{I}+\mathbf{L}_{i}\mathbf{U}_{i}|-\text{Tr}\{\mathbf{Q}_{i}^{\mathrm{H}}(\mathbf{L}_{i}^{-1}+\mathbf{U}_{i})^{-1}\mathbf{Q}_{i}\}$.

The update of $\mathbf{L}_{i}$ only depends on $\mathcal{L}(i)$. To optimize $\mathcal{L}(\boldsymbol{\Theta})$, we set $\frac{\partial \mathcal{L}(i)}{\partial \mathbf{L}_{i}}=\mathbf{O}$. The update rule of $\mathbf{L}_{i}$ is formulated as
\begin{equation}
	\mathbf{L}_{i}=\mathbf{U}_{i}^{-1}(\frac{\mathbf{Q}_{i}\mathbf{Q}_{i}^{\mathrm{H}}}{L}-\mathbf{U}_{i})\mathbf{U}_{i}^{-1}.
	\label{L1}
\end{equation}

Thus, the intra-block correlation $\gamma_{i}$ is given by \cite{liu2014energy}
\begin{equation}
	\small
	\gamma_{i}=\frac{1}{2}\text{Tr}(\mathbf{L}_{i}).
	\label{gamma}
\end{equation}

Update the correlation structure matrix $\mathbf{G}_{i}={\mathbf{L}_{i}}/\gamma_{i}$. 

The convergence condition of the algorithm is satisfied when the normalized change rate $\epsilon=\frac{\Vert \boldsymbol{\gamma}^{\text{new}} - \boldsymbol{\gamma}^{\text{old}} \Vert_2}{\Vert \boldsymbol{\gamma}^{\text{old}} \Vert_2} $ falls below the predefined threshold $\epsilon_{min}$, or when the maximum number of iterations is reached.

\subsection{Block structure factors and noise value estimation.}
For block-sparse models of NC signals, further constraints on \( \mathbf{G}_i \) can enhance algorithm performance due to the fixed block size \cite{10328457}. A first-order autoregressive (AR) model is used in \(\mathbf{G}_{i}\) to efficiently capture intra-block correlations along the main and sub-diagonals while reducing computation \cite{6415293},
\begin{equation}
	\mathbf{G}_{i}=\text{Toeplitz}([1,r]),
	\label{G_toeplitz}
\end{equation}
the AR factor \( r \) is defined as $r \triangleq \frac{m_{1}}{m_{0}}$, where $m_{1}$ and $m_{0}$ represents the mean value of the elements on the main diagonal and the sub-diagonal of $\mathbf{G}_{i}$ respectively.

The noise parameter $\beta^{-1}$ can be estimated using the following formula
\begin{equation}
	\beta = \frac{M}{\text{Tr}[\boldsymbol{\Sigma}\mathbf{V}^{\mathrm{H}}\mathbf{V}]+\Vert \mathbf{Y}-\mathbf{V}\mathbf{X} \Vert_\mathrm{F}^{2}}.
\end{equation}

However, in practical applications, updating \(\beta\) can lead to instability in signal reconstruction performance \cite{liu2014energy}. Therefore, \(\beta\) is considered a constant set manually as
$\beta^{-1}=0.01\Vert \mathbf{Y} \Vert_\mathrm{F}^{2}$.

\subsection{DOA and NC phases estimation}
The hyperparameter \(\boldsymbol{\gamma}\) from (\ref{gamma}) determines the sparse signal's power distribution, enabling the reconstruction of \(\hat{\mathbf{X}}\), where \(\hat{\mathbf{X}}_{i} \in \mathbb{C}^{2\times L}\) represents the \(n\)-th block. The spectral function is then given as
\begin{equation}
	\mathbf{P}(\tilde{\theta}_n)=\frac{1}{L}(\sum_{l=1}^{L}|\hat{\mathbf{X}}_{i,1,l}|+\sum_{l=1}^{L}|\hat{\mathbf{X}}_{i,2,l}|),
	\label{P}
\end{equation}
where \(\hat{\mathbf{X}}_{i,1,l}\) and \(\hat{\mathbf{X}}_{i,2,l}\) represent the \(l\)-th sample data in the first and second rows of the \(n\)-th block, respectively. The \(K\) largest spectral peaks, identified via a spectral peak search, correspond to the desired DOA values.

The proposed algorithm estimates the DOA and recovers the \(k\)-th NC phase from the reconstructed signal \(\hat{\mathbf{X}}\), with the estimation formula given by  
\begin{equation}
	\hat{\phi}_{k}=\frac{\sum_{l=1}^{L}\mathrm{arctan}(\mathrm{Im}(\hat{\mathbf{X}}_{k,1,l})/\mathrm{Re}(\hat{\mathbf{X}}_{k,1,l}))}{L},
	\label{phi}
\end{equation}
where $\hat{\mathbf{X}}_{k}$ represents the block signal corresponding to the $k$-th spectral peak.
The algorithm flow is presented in Algorithm \ref{algorithm1}.
\begin{algorithm}
	\scriptsize
	\caption{\enskip NC-BSBL}\label{alg1}
	\begin{algorithmic}[1]
		\STATE \textbf{Input:} \(\mathbf{Y}\), \(\overline{\mathbf{B}}\), \(\mathbf{J}\), obtain \(\mathbf{V}\) from (\ref{V}).
		\STATE \textbf{Initialization:} Initialize \(\mathbf{G}_i=\mathbf{I}\), \(\mathbf{L}_i=\mathbf{O}\), \(\mathbf{U}_i=\beta \mathbf{V}^\mathrm{H} \mathbf{V}\), \(\mathbf{Q}_i=\beta \mathbf{V}^\mathrm{H} \mathbf{Y}\), set parameters \(\text{Maxiter}=500\) and \(\epsilon_{\text{min}}=10^{-4}\).
		\WHILE{\(\epsilon>\epsilon_{\text{min}}\) \textbf{and} iter $<$ \text{Maxiter}}
		\STATE Update \(\mathbf{L}_i\), \(\gamma_i\), \(\mathbf{G}_i\) (\ref{L1}), (\ref{gamma}), (\ref{G_toeplitz}).
		\STATE Compute \(\Delta \mathcal{L}(i)=\mathbf{L}(\gamma_{i}\mathbf{G}_{i})-\mathcal{L}(\mathbf{L}_{i})\) for all blocks and select \(\hat{i}\)-th block suck that \(\Delta \mathcal{L}(\hat{i}) = \min \{\mathcal{L}(i)\}\).
		\STATE Update \(\boldsymbol{\mu}\), \(\boldsymbol{\Sigma}\) based on (\ref{mu}), (\ref{sigma}).
		\STATE iter = iter + 1.
		\ENDWHILE
		\STATE Compute spectral peak \(\mathbf{P}(\tilde{\theta}_n)\) (\ref{P}) and NC phases (\ref{phi}).
		\STATE \textbf{Output:} DOA parameters \(\hat{\theta}\) and NC phases \(\hat{\phi}\).
	\end{algorithmic}
	\label{algorithm1}
\end{algorithm}

\section{Numerical Simulation}
\label{experiments}
This section compares the proposed algorithm with NC-PM \cite{chen2015doa}, NC-ESPRIT \cite{wang2017non}, SBL \cite{7536146}, and its variant adapted for NC signals (NC-SBL), which assumes known NC phase information incorporated into the overcomplete dictionary $\mathcal{G}$.

The experimental setup includes \( M = 12 \) array elements, \( L = 20 \) snapshots, an overcomplete dictionary \(\mathcal{G} = [-40:0.1:40]^\circ\) with \( N = 801 \) grid points, \(\epsilon_{\text{min}} = 10^{-4}\), and a maximum of 500 iterations.  NC-SBL requires identical NC phases $\boldsymbol{\Psi}_{1}= [\pi/12, \pi/12]$, while NC-BSBL adapts for arbitrary NC phases $\boldsymbol{\Psi}_{2}=[\pi/12, \pi/4]$. $\boldsymbol{\theta}=[10^\circ,30^\circ]$.

We employ the root mean square error (RMSE) as a metric to evaluate the accuracy of DOA and NC phases estimation, which is defined as \cite{10124337}
\begin{equation}
	\small
	\text{RMSE}= \sqrt{\frac{1}{K}\frac{1}{T}\sum_{k=1}^{K}\sum_{t=1}^{T}(\hat{\xi}_{t,k}-\xi_{k})^2}.
\end{equation}
where \(\hat{\xi}_{t,k}\) denotes the estimated DOA (NC phases) parameter for the $k$-th source in the $t$-th trial, while \(\xi_{k}\) represents its true value of the $k$-th source. The number of trials for the Monte Carlo simulation is set to \( T = 500 \).
The Cramér-Rao Bound (CRB) \cite{7468573} is utilized to assess the algorithm's performance. 

Table \ref{table:cpu_time} presents the average CPU time required for 500 runs of the algorithm under $\text{SNR}=10\text{dB}$. It can be observed that the NC-BSBL algorithm significantly reduces the computation time due to the incorporation of the FMLM method.
\vspace{-3mm}
%
\begin{table}[htbp]
	\small
	\centering
	\caption{Average CPU time for 500 runs of algorithms.}
	\begin{tabular}{lcc}
		\toprule
		Algorithm & Average CPU Time (Seconds) \\
		\midrule
		NC-BSBL (FMLM)  & 0.1165  \\
		SBL (No FMLM) & 2.1516  \\
		NC-SBL (No FMLM) & 2.3236  \\
		\bottomrule
	\end{tabular}
	\label{table:cpu_time}
\end{table}

\begin{figure}[!ht]
	\subfloat[]{\includegraphics[width=0.165\textwidth]{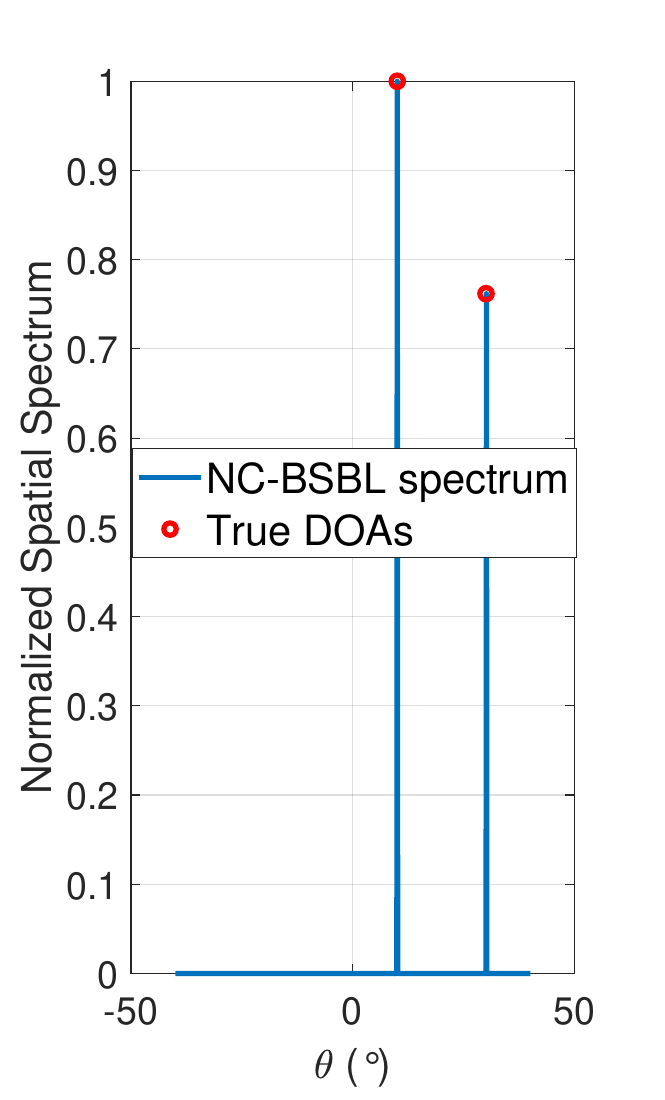}} 
	\subfloat[]{\includegraphics[width=0.165\textwidth]{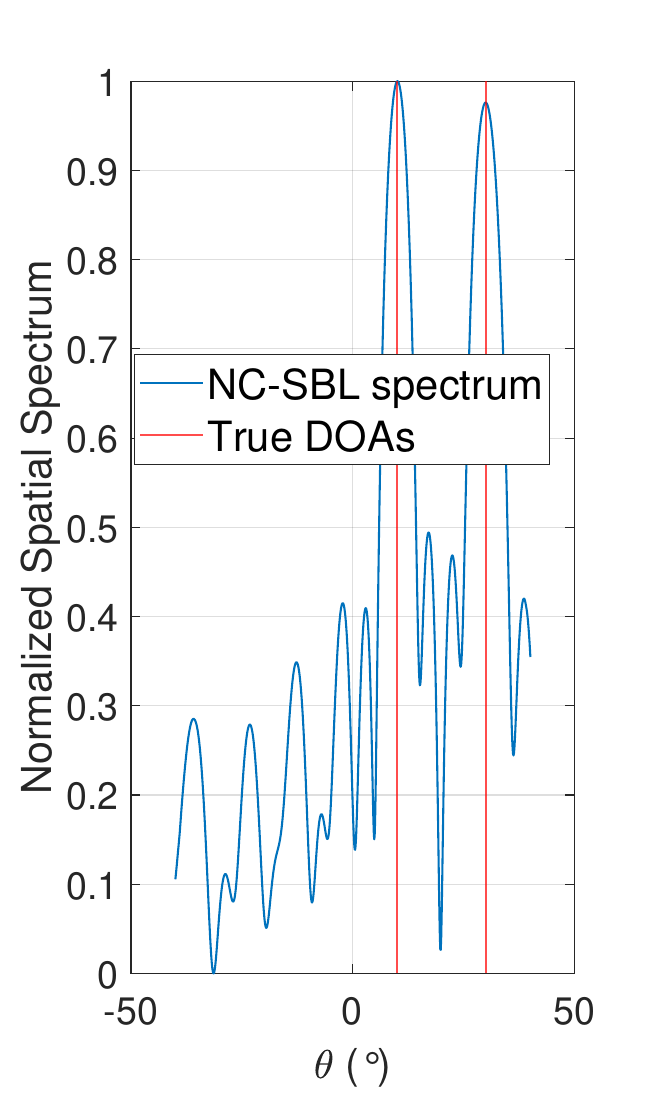}} 
	\subfloat[]{\includegraphics[width=0.165\textwidth]{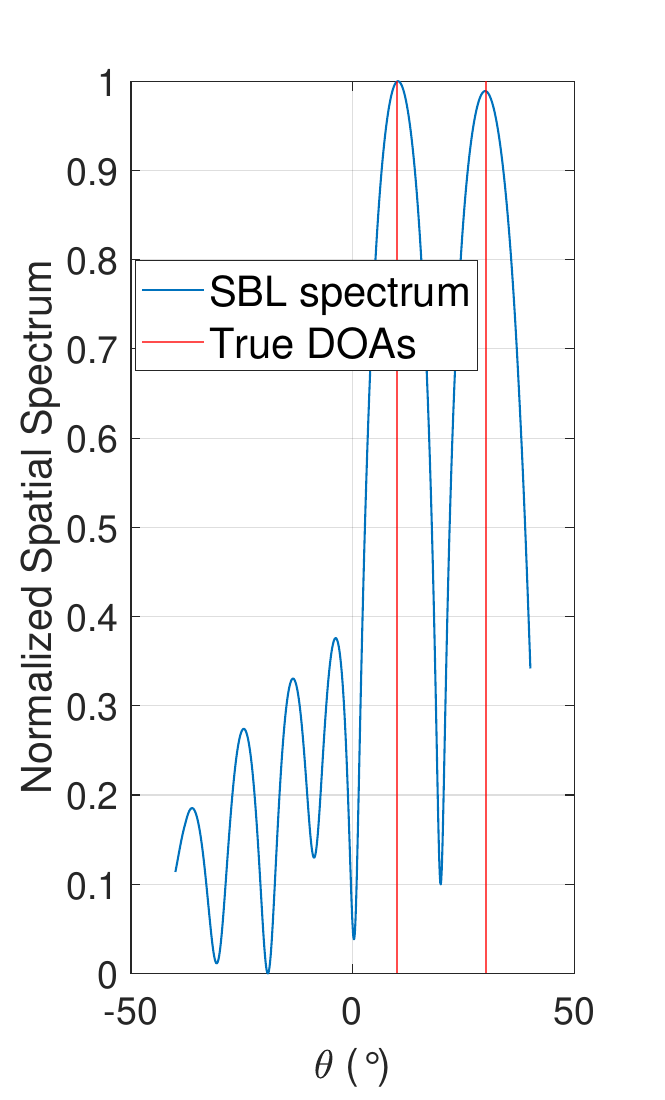}} 
	\caption{Spectral peaks for two NC signals, $\boldsymbol{\theta}=[10^\circ, 30^\circ]$, $\boldsymbol{\Psi}_{1}=[\pi/12,\pi/12]$ (known for NC-SBL), $\boldsymbol{\Psi}_{2}=[\pi/12,\pi/4]$ (unknown for NC-BSBL), $M=12$, SNR = 10 dB, $L=20$. (a) NC-BSBL. (b) NC-SBL. (c) SBL.}
	\label{spectrum}
\end{figure}

Fig. \ref{spectrum} presents the spectral peaks of NC-BSBL, NC-SBL, and SBL. It can be observed that NC-BSBL produces fewer spurious peaks, with sharper peaks at the correct DOAs, indicating that the proposed algorithm achieves more accurate estimates compared to the others.

\begin{figure}[!ht]
	\centerline{\includegraphics[width=0.62\columnwidth]{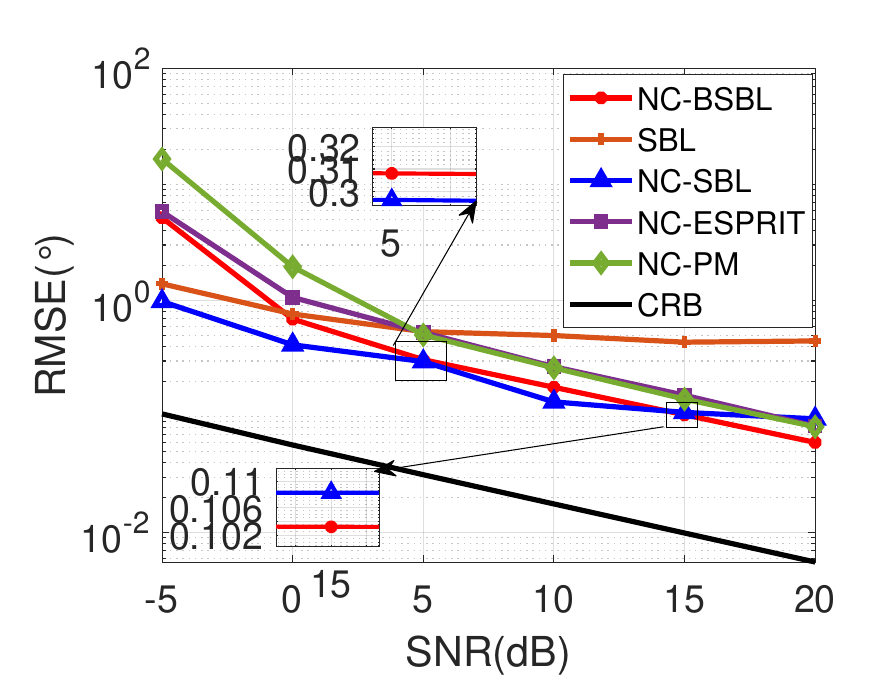}}
	\caption{The RMSE curves as a function of SNR with \( M = 12 \), \( L = 20 \), \(\boldsymbol{\theta} = [10^\circ, 30^\circ]\), and \(\boldsymbol{\Psi}_{1} = [\pi/12, \pi/12]\) (known for NC-SBL), \(\boldsymbol{\Psi}_{2} = [\pi/12, \pi/4]\) (unknown for NC-BSBL), $T=500$.}
	\label{rmse_doa}
\end{figure}

Fig. \ref{rmse_doa} illustrates the estimation error of various algorithms across SNRs from -5dB to 20dB. While noncircularity increases signal dimensionality and amplifies noise, impacting FMLM's basis selection and performance \cite{9540245}, NC-BSBL shows higher errors at low SNRs. However, at high SNRs (e.g., above 15dB), it surpasses SBL and NC-SBL by leveraging NC properties and intra-block correlations for enhanced accuracy.

\begin{figure}[!ht]
	\centerline{\includegraphics[width=0.62\columnwidth]{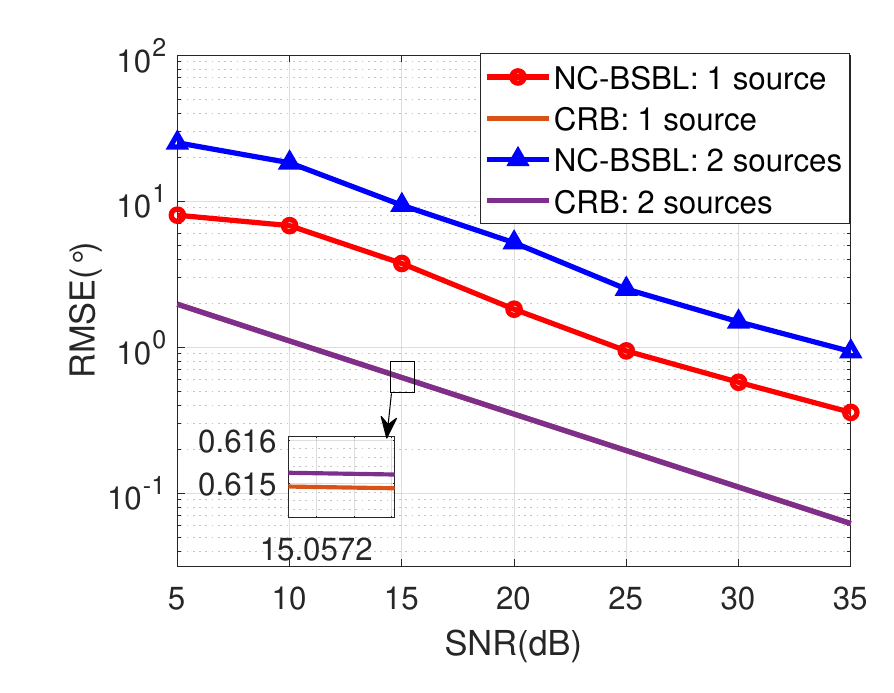}}
	\caption{The curve of RMSE versus SNR for NC phases, with $M=12$, $L=20$, $T=500$.}
	\label{rmse_fai}
\end{figure}

Fig. \ref{rmse_fai} illustrates the RMSE of NC-BSBL for NC phases recovery under single-source (\(\theta = 10^\circ\), \(\phi = \pi/4\)) and two-source (\(\boldsymbol{\theta} = [10^\circ, 30^\circ]\), \(\boldsymbol{\Psi} = [\pi/12, \pi/4]\)) scenarios. As SNR increases, improved source reconstruction enhances NC phase recovery, validating the algorithm's effectiveness in estimating unknown and distinct NC phases.
 
\section{Conclusion}
This letter has proposed a DOA estimation algorithm for NC signals in the BSBL framework, leveraging noncircularity and block sparsity without prior NC phase. The permutation matrix reorganizes signal components to capture intra-block information, improving recovery. Future work will refine block constraints via NC conjugate symmetry and integrate off-grid techniques for greater precision.  

\bibliographystyle{IEEEtran}
\bibliography{IEEEabrv,ref.bib}
\end{document}